\documentstyle[aps,prl,multicol,psfig]{revtex}\def\tc{yy}
\def\inodot{\char'020}
\def\<{\langle}
\def\>{\rangle}
\def\({\left (}
\def\){\right )}
\def\[{\left [}
\def\]{\right ]}

\def\i{{\rm i}}
\def\e{{\rm e}}

\begin{document}
\title{Classical and Quantum Hamiltonian Ratchets}
\author{Holger Schanz,$^{1}$ Marc-Felix Otto,$^{1}$ Roland Ketzmerick,$^{1}$
and Thomas Dittrich$^{2}$}
\address{$^{1}$Max-Planck-Institut f\"ur Str\"omungsforschung und Institut
f{\"u}r Nichtlineare Dynamik\\ der Universit{\"a}t G{\"o}ttingen,
Bunsenstra\ss e 10, 37073 G\"ottingen, Germany}
\address{$^{2}$Departamento de F\'\inodot sica, Universidad Nacional,
Santaf\'e de Bogot\'a, Colombia}
\date{June 10, 2001}
\maketitle
\begin{abstract}
We explain the mechanism leading to directed chaotic transport in Hamiltonian
systems with spatial and temporal periodicity.  We show that a mixed phase
space comprising both regular and chaotic motion is required and derive a
classical sum rule which allows to predict the chaotic transport velocity from
properties of regular phase-space components. Transport in quantum
Hamiltonian ratchets arises by the same mechanism as long as uncertainty
allows to resolve the classical phase-space structure. We derive a quantum
sum rule analogous to the classical one, based on the relation between quantum
transport and band structure.
\end{abstract}
\if\tc\begin{multicols}{2}\fi 
Stimulated by the biological task of explaining the functioning of molecular
motors, the study of ratchets \cite{Feynman} has widened to a general
exploration of ``self-organized'' transport, i.e., transport without external
bias, in nonlinear systems \cite{JAP97}. Along with this process, there has
been a tendency to reduce the models under investigation from realistic
biophysical machinery to the minimalist systems customary in nonlinear
dynamics. External noise, for example, which originally served to account for
the fluctuating environment of molecular motors, has been replaced by
deterministic chaos. This required to include inertia terms in the equations
of motion, thus leaving the regime of overdamped dynamics and leading to
deterministic inertia ratchets with dissipation \cite{JKH96,Mat00}. It is then
a consequent but radical step to abandon friction altogether. Indeed,
transport in \emph{Hamiltonian ratchets} was observed numerically if all
symmetries were broken that generate to each trajectory a countermoving
partner \cite{FYZ00,GH00}. 

As a parallel development, the desire to realize ratchets in artificial,
nanostructured electronic systems, required to consider quantum effects
\cite{RGH97,GH00}. \emph{Quantum Hamiltonian ratchets}, however, have been
studied only in the framework of one-band systems where no transport occurs
\cite{GH00}.

In this paper we explain how a Hamiltonian ratchet works. We rely on methods
which---although well established in studies of deterministic dynamics---have
never before been applied to ratchets.  We derive a classical and an analogous
quantum sum rule for transport allowing the following conclusions: (i)
Directed transport is a property associated with individual invariant sets of
the dynamics. A necessary condition for non-zero transport is a mixed phase
space with coexisting regular and chaotic regions.  (ii) Transport in {\em
chaotic} regions can be described quantitatively by using topological and
further properties of adjacent {\em regular} regions only.  (iii) Quantum
transport persists for all times and approaches the classical transport when
$\hbar$ is small compared to the major invariant sets of the classical phase
space.

\def\figI{
\begin{figure}[tb]
 \centerline{\psfig{figure=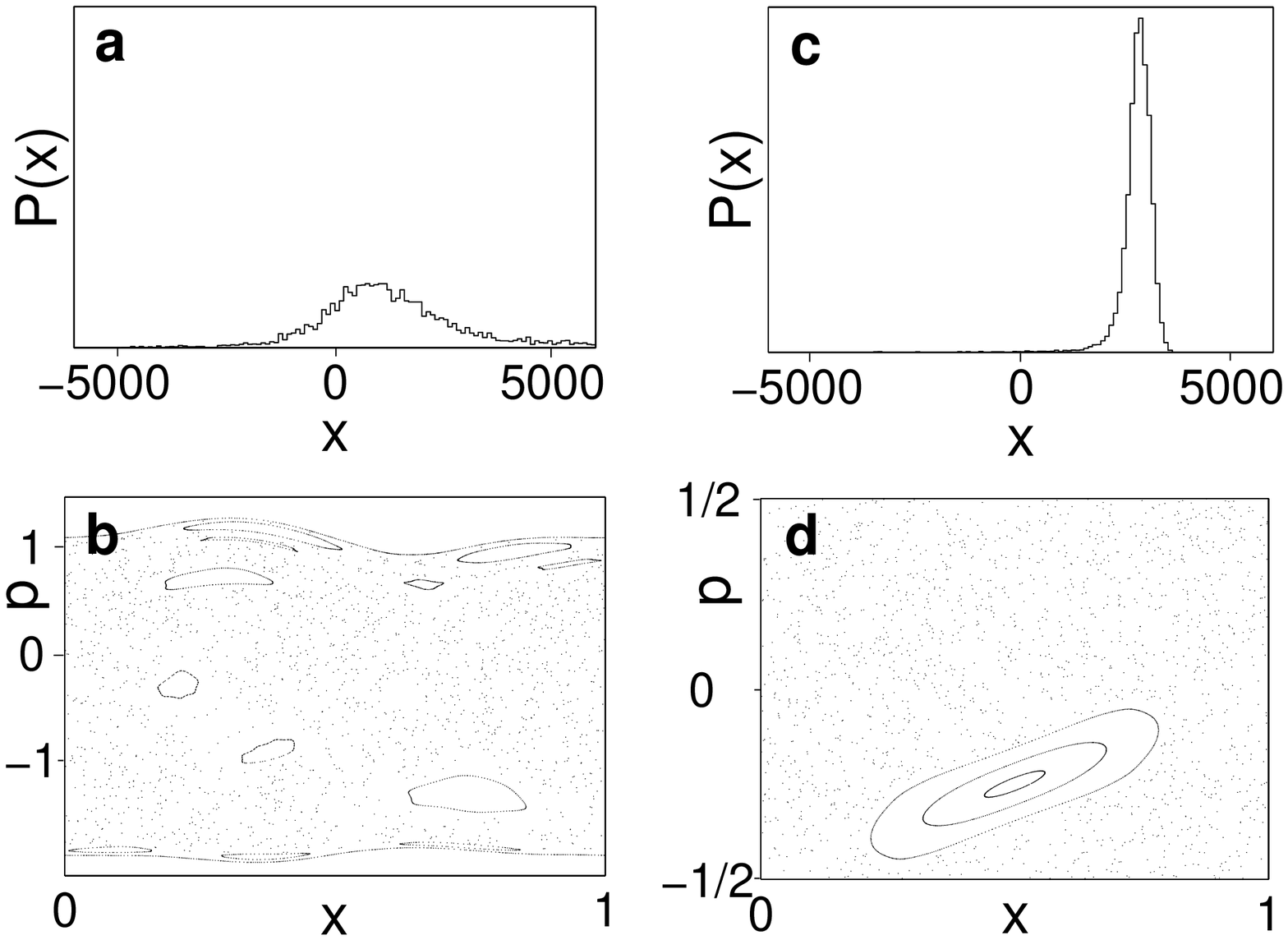,width=\fw}} \caption{\label{figI} (a)
  Spatial distribution $P(x)$ of a continuously driven system
  \protect\cite{FYZ-Parameters} after $20,000$ time periods showing the
  directed transport in a Hamiltonian ratchet. Initially, $10^{4}$
  trajectories were started at random on the line $p=0$, $x\in [0,1)$ in the
  chaotic sea.  (b) Poincar\'e section $p$ vs $x$ of a unit cell at integer
  times showing the main chaotic sea, the upper and lower limiting KAM-tori,
  and the major embedded regular islands.  (c,d) As (a,b), but for the kicked
  Hamiltonian (\protect\ref{KickH}) showing a much more pronounced directed
  transport.}
\end{figure}
}
\if\tc{\def\fw{78mm}\figI}\fi

We consider a Hamiltonian of the form $H(x,p,t)=T(p)+V(x,t)$, where $T(p)$ is
the kinetic energy. The force $-V'$ is periodic in space and time,
$V'(x+1,t)=V'(x,t+1)=V'(x,t)$, and has zero mean $\int_{0}^{1}{\rm
d}t\,\int_{0}^{1}{\rm d}x\,V'(x,t)=0$. Usually directed transport is
demonstrated by following selected trajectories over very long times
\cite{FYZ00,GH00} or an ensemble of trajectories which generates spatial
distributions as shown in Fig.~\ref{figI}a,c. While this is easily implemented
numerically, it gives no clue about the origin of the transport (but see
Ref.~\cite{DF01}). Instead, we shall exploit the periodicity of the dynamics
with respect to space and time and analyze transport in terms of the invariant
sets of phase space, \emph{reduced to the spatio-temporal unit cell} $x,t\in
[0,1)$.  For any finite invariant set $M$ we define ballistic transport as
phase-space volume times average velocity expressed as
\begin{equation}\label{transport}
\tau_{M}=
\int_{0}^{1}{\rm d}t\,
\int_{0}^{1}{{\rm d}x}\,
\int_{-\infty}^{+\infty}{\rm d}p\,
\chi_{M}(x,p,t)\,{\partial H\over\partial p}\,,
\end{equation}
where $\chi_{M}$ is the characteristic function of $M$. Transport is additive
for the union of two or more disjunct invariant sets, i.e., for
$M=\bigcup_{i} M_{i}$, with $M_i \cap M_j = \emptyset$ for all $i \neq j$, we
have
\begin{equation}\label{SumRule}
\tau_{M}=\sum_{i}\tau_{M_{i}}\,.
\end{equation}
This sum rule for Hamiltonian transport has far-reaching consequences to be
discussed in the following.

For a generic Hamiltonian system, phase space is mixed and comprises an
infinite number of minimal invariant sets of different types. For the sake of
definiteness we will restrict the following discussion to the most interesting
case of a chaotic region containing embedded regular islands
(Fig.~\ref{figI}b). In any of these invariant sets the time-averaged velocity
$v_i$ is the same for almost all initial conditions (assuming ergodicity for
chaotic components).  Hence, for a chaotic region, $\tau_{\rm ch}=A_{\rm
ch}\,v_{\rm ch}$ with $A_{\rm ch}$ denoting its area in a stroboscopic
Poincar\'{e} section. For an embedded island we have $\tau_i=A_i\,v_i$ where
$A_i$ includes the areas of the narrow chaotic layers inside the island and of
the infinite hierarchy of island chains surrounding it because all these
invariant sets share the same mean velocity. This velocity $v_{i}$ is
identical to the rational winding number $w_{i}=x_{i}/t_{i}$ of the stable
fixed point at the center of the island. In extended phase space, this
corresponds to a shift of the island by $x_{i}$ spatial after $t_{i}$ time
periods. Typically, the chaotic set is bounded from above and below by two
non-contractible KAM-tori $p_{a,b}(x,t)$ enclosing the spatial unit
cell. Treating the phase-space region in between as the global invariant set
$M$ appearing on the l.h.s.\ of Eq.~(\ref{SumRule}), its transport $\tau_{M}$
is obtained from Eq.~(\ref{transport}) as $\<T\>_{{a}}-\<T\>_{{b}}$ with
$\<T\>_{{a,b}}=\int_{0}^{1}{\rm d}t\,\int_{0}^{1}{\rm d}x\,T(p_{a,b}(x,t))$
denoting averages of the kinetic energy over the tori.  Using the sum rule
(\ref{SumRule}) we can now express transport of the chaotic region in terms of
its adjoining regular components (KAM-tori and islands) as
\begin{equation}\label{ChTr}
A_{\rm ch}v_{\rm ch}=\<T\>_{{a}}-\<T\>_{{b}}-\sum_{i}A_i\,w_{i}\,.
\end{equation}
This is our main result on classical transport in Hamiltonian ratchets. Not
only does Eq.~(\ref{ChTr}) provide an efficient method to determine the
chaotic drift velocity, it also expresses the simple principle generating
directed ballistic motion: Decomposing phase space into different invariant
sets, these will in general have average velocities different from each other
and also different from zero but related by the sum rule
(\ref{SumRule}). Therefore a necessary condition for directed chaotic
transport in Hamiltonian ratchets is \emph{a mixed phase space}.

L\'evy flights \cite{Theo} are a characteristic feature of chaotic motion in a
generic mixed phase space and indeed they were observed in Hamiltonian
ratchets \cite{FYZ00,GH00}. They reflect the slow exchange between subsets of
a chaotic region, separated by leaky barriers \cite{Theo}. As these subsets
are not invariant, their contributions to Eq.~(\ref{SumRule}) are contained in
the contribution of the chaotic invariant set. L\'evy flights lead to
power-law tails in spatial distributions. For example, the asymmetric shapes
of the peaks visible in Figs.~\ref{figI}a,c can be attributed to these tails.
Notably, in Fig.~\ref{figI}c one clearly sees a mean transport to the right
although the same data shows no indication of a power-law tail in this
direction. We stress that the sum rule allows to predict the mean velocity of
chaotic trajectories without any reference to such details of the chaotic
dynamics. This suggests that L\'evy flights are not a necessary element of the
mechanism of chaotic transport in Hamiltonian ratchets.

In Ref.~\cite{FYZ00} it was shown that a necessary condition for directed
transport is the breaking of all symmetries which to each trajectory generate
a countermoving partner.  For a chaotic set invariant under such a
symmetry, this is in agreement with Eq.~(\ref{ChTr}) because then the r.h.s.\
vanishes identically. However, chaotic sets can also occur as
symmetry-related pairs transporting in opposite directions.  Moreover, if
phase space cannot be decomposed into invariant subsets, e.g., for an
ergodic system, there cannot be transport even with all symmetries broken.

Up to now we have only considered transport of invariant sets of the unit
cell.  For an arbitrary initial distribution transport is determined by
projection onto these invariant sets \cite{D+00}. Therefore, the location of
an initial distribution \emph{within an invariant set} is irrelevant. This
applies also to the location within the temporal unit cell, i.e., to the
question of phase dependence discussed in \cite{YFR00}. In particular, in case
that the plane $p=0$, $0\le x,t\le 1$ is completely within the chaotic
invariant set, any initial condition restricted to this plane will result in
the same average transport. We now understand how a Hamiltonian ratchet makes
particles initially at rest ($p=0$) move with a predetermined mean velocity
as, e.g., in Fig.~1a.

We have checked Eq.~(\ref{ChTr}) numerically for a continuously driven system
\cite{FYZ-Parameters}. We determined the areas $A_{i}$ and winding numbers
$w_{i}$ for the regular islands shown in the Poincar\'e section of
Fig.~\ref{figI}b as well as $\<T_{a}\>$ and $\<T_{b}\>$ for the limiting
KAM-tori, yielding $v_{\rm ch}=0.092\pm 0.011$. The error estimate includes
the uncertainty in the location of the bounding KAM-tori and the contribution
from neglected small islands. The result is in agreement with the value
$v_{\rm ch}=0.082\pm 0.002$ determined with much more computational effort
from the spatial distribution of $10^{4}$ trajectories, started with $p=0$
(Fig.~\ref{figI}a).

As a minimal model for directed chaotic transport in Hamiltonian ratchets,
we propose a kicked Hamiltonian
\begin{equation}\label{KickH}
H(x,p,t)=T(p)+V(x)\,\sum_{n}\delta(t-n).
\end{equation}
It reduces the dynamics to a map for position and momentum $x_{n+1} =
x_n+T'(p_{n})$, $p_{n+1} = p_n-V'(x_{n+1})$, just after the kick. As an
example we take a symmetric potential $V(x)=(x\,{\rm mod}\,1\,-1/2)^{2}/2$ and an
asymmetric kinetic energy $T(p)=|p|+3\sin(2\pi p)/(4\pi^{2})$.
We consider the dynamics on a cylinder with transport along the $x$-axis and
$p\in [-1/2,+1/2)$ being a periodic variable. Fig.~\ref{figI}d shows the
Poincar\'e section for one unit cell. There are only two major invariant
sets---a chaotic sea and a regular island centered around a periodic orbit
with winding number $w_{\rm reg}=-1$. According to Eq.~(\ref{transport}),
transport of the full phase space vanishes identically because of the periodic
momentum variable.  Applying the sum rule (\ref{SumRule}) the contributions to
transport from the two invariant sets cancel exactly,
\begin{equation}\label{krsum}
A_{\rm ch}\,v_{\rm ch}+A_{\rm reg}\,w_{\rm reg}=0\,.
\end{equation}
We find the transport velocity of the chaotic component as $v_{\rm ch}=f_{\rm
reg}/(1-f_{\rm reg})$, where $f_{\rm reg}=A_{\rm reg}/(A_{\rm reg}+A_{\rm
ch})$ denotes the relative area of the regular island. From Fig.~\ref{figI}d,
$f_{\rm reg}=0.117\pm 0.001$, thus $v_{\rm ch}=0.133\pm 0.001$ in agreement
with $v_{\rm ch}=0.1344\pm 0.0003$ from the spatial distribution of
Fig.~\ref{figI}c.
\def\figII{
\begin{figure}
 \centerline{ \psfig{figure=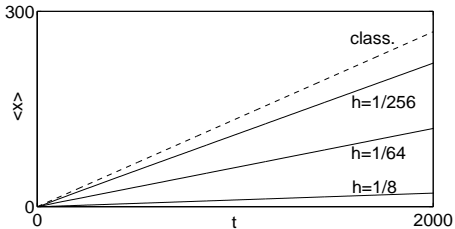,width=\fw}\hspace*{3mm} }
  \caption{\label{figII} Mean position vs time for a wavepacket in system
  (\protect\ref{KickH}) initialized as the momentum eigenstate with $p=0$ for
  various values of $h$ (full lines). For decreasing $h$ the classical
  prediction $v_{\rm ch}\,t$ (dashed line) is approached.}
\end{figure}
}
\if\tc{\def\fw{60mm}\figII}\fi
In order to extend our concept of directed transport in Hamiltonian systems to
quantum ratchets, we first demonstrate by a numerical example that quantum
Hamiltonian ratchets can work. Fig.~\ref{figII} shows that the average
velocity of a wavepacket initialized in the chaotic sea varies between $0$ for
large and the classical value $v_{\rm ch}$ for small values of $\hbar$. We
explain this behavior in the following.

In analogy with our approach to classical transport, we consider the
invariants of the quantum dynamics, the stationary states of the
time-evolution operator over one period, i.e., $\hat U=\e^{-\i \hat
V/\hbar}\e^{-\i \hat T/\hbar}$ for the kicked Hamiltonian
Eq.~(\ref{KickH}). They satisfy $\hat U|\phi_{\alpha,k}\>=\exp(-2\pi
\i\epsilon_{\alpha}(k))\,|\phi_{\alpha,k}\>$, with the quasienergy
$\epsilon_{\alpha}(k)\in[0,1)$ \cite{Sam73}. Similarly, spatial periodicity
implies $\phi_{\alpha,k}(x+1,t)=\exp(2\pi\i k)\,\phi_{\alpha,k}(x,t)$ with
quasimomentum $k\in[0,1)$ where $h$ is chosen rational for systems periodic in
$p$. Quantum transport is related to the expectation values in the stationary
states $\overline v_{\alpha,k}\equiv\langle\langle{\phi_{\alpha,k}|\hat
v|\phi_{\alpha,k}}\rangle\rangle$ of the velocity operator $\hat v=\hat
T'(\hat p)$, where $\langle\langle\cdot\rangle\rangle=\int_{0}^{1}{\rm
d}x\int_{0}^{1}{\rm d}t(\cdot)$. Using a generalization of the Hellman-Feynman
theorem to time-periodic systems \cite{Sam73}, we express velocities by
\emph{band slopes} as
\begin{eqnarray}\label{hell}
\overline v_{\alpha,k}={\rm d}\epsilon_{\alpha}(k)/{\rm d}k\,.
\end{eqnarray}
This allows to discuss quantum transport in terms of spectral properties.
Examples for quasienergy band spectra are shown in Fig.~\ref{figIII} together
with the corresponding velocity distributions. The semiclassical regime is
characterized by the existence of two different types of bands and
corresponding eigenstates \cite{KMG94,K+00}: Bands pertaining to regular
states appear as straight lines in the spectrum, while the chaotic bands show
oscillations and wide avoided crossings among themselves. Associating the
terms chaotic and regular with the bands is supported by the Husimi
representations of the corresponding eigenfunctions (insets in
Fig.~\ref{figIII}a).  The new aspect introduced into this picture by directed
chaotic transport is the \emph{overall slope of the chaotic bands}.
\def\figIII{
\begin{figure}
 \centerline{\psfig{figure=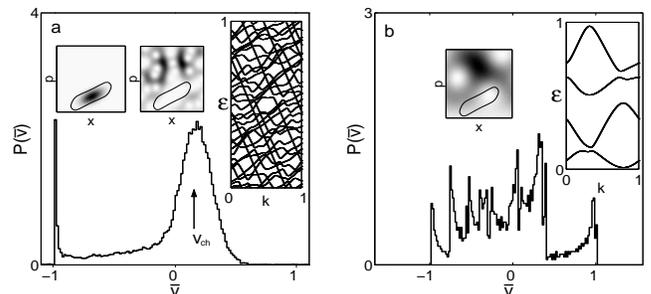,width=\fw}}
 \caption{\label{figIII}
  Distribution of quantum velocities $\overline v$ obtained according to
  Eq.~(\protect\ref{hell}) for (a) $h=1/128$ and (b) $h=1/4$. The right insets
  show band spectra for $h=1/32$ and $h=1/4$, respectively. The smaller insets
  are Husimi representations of characteristic wave functions together with
  the border of the classical regular island. In (a) the regular and the
  chaotic wave functions can be associated with the two peaks of $P(\overline
  v)$ centered around $v_{\rm reg}=-1$ and $v_{\rm ch}$.}
\end{figure}
}
\if\tc{\def\fw{85mm}\figIII}\fi

Only on a coarse quasienergy scale, the two sets of bands appear to cross.  On
a sufficiently fine scale, all crossings are avoided.  Consequently the actual
bands change their character between regular and chaotic at each of the narrow
crossings and have {\em no} overall slope.  Switching from the latter
(``adiabatic'') to the former (``diabatic'') viewpoint is a well-controlled
procedure \cite{KMG94}.  Formally, this behavior of the bands can be described
in terms of their winding number (average slope) with respect to the periodic
$(\epsilon,k)$-space: In the adiabatic as well as in the diabatic case, all
quasienergy bands must close after an integer number of periods in the
$\epsilon$ and $k$ directions, so that their winding number $\overline w$ must
be rational. Clearly, in the adiabatic case, all winding numbers are
zero. Going from the adiabatic to the diabatic case amounts to a mere
reconnection of bands at the crossings, preserving the sum of winding
numbers. Thus it must be zero also in the diabatic representation,
\begin{equation}\label{qsum}
\sum_\alpha \overline w_\alpha^{\rm (ch)} + \sum_\alpha \overline
w_\alpha^{\rm (reg)}=0\,.
\end{equation}
This is the quantum-mechanical analogue of the classical sum rule
(\ref{krsum}). Because of the localization of the regular states on tori
inside the regular island, the winding number of the regular bands in
$(\epsilon,k)$-space is in the semiclassical limit identical to the winding
number in $(x,t)$-space of the central periodic orbit, i.e., $\overline
w_\alpha^{\rm (reg)}=w_{\rm reg}$.  Moreover, in this limit, the fractions of
regular and chaotic bands correspond to the relative phase-space volumes
$f_{\rm reg}$ and $1-f_{\rm reg}$, respectively. We therefore obtain from
Eq.~(\ref{qsum}) the mean slope of the chaotic bands, $\overline w^{\rm
(ch)}=f_{\rm reg}/(1-f_{\rm reg})=v_{\rm ch}$, as the classical drift
velocity. This is confirmed in Fig.~\ref{figIII}a. 

The asymptotic quantum transport velocity for a given initial wavepacket
$|\psi\rangle$ is an average of band slopes weighted with the overlaps
$|\langle\psi|\phi_{\alpha,k}\rangle|^2$.
We can now explain our observations in Fig.~\ref{figII}: For $\hbar\ll A_{\rm
reg}$, an initial wavepacket prepared in the chaotic region of a single unit
cell of the extended system is a superposition of chaotic eigenfunctions from
the entire band spectrum.  Consequently, its drift velocity is given by the
mean slope of the chaotic bands and thus by the classical value $v_{\rm
ch}$. In contrast, for $\hbar\gg A_{\rm reg}$, there are no states restricted
to the regular or the chaotic set and hence quantum transport does not
correspond to classical transport in this regime.
\def\figIV{
\begin{figure}
 \centerline{\psfig{figure=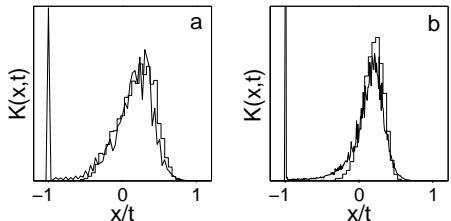,width=\fw}} \caption{\label{figIV} Form
 factor $K(x,t)$ for $h=1/128$ (thick line) at (a) $t=30$ and (b) $t=111$, the
 Heisenberg time of the chaotic component.  The two distinct peaks centered
 around $v_{\rm reg}$ and $v_{\rm ch}$ are the fingerprints of directed
 transport. The time-dependent width of the chaotic peak follows the classical
 distribution $P^{\rm (ch)}(x,t)$ of chaotic trajectories (thin line).}
\end{figure}
}
\if\tc{\def\fw{60mm}\figIV}\fi

Our analysis based on the winding numbers can be applied to predict the mean
quantum transport in the semiclassical regime from the classical value. The
band spectra, however, encode more detailed information about quantum
transport. It can be extracted by a double Fourier transform $\epsilon \to t$,
$k \to x$ (discrete position in units of the spatial period) and subsequent
squaring of the spectral density, translating two-point correlations in the
bands into the entire time evolution of the spatial distribution on the scale
of the spatial period. A formal definition of the resulting \emph{generalized
form factor} $K(x,t)$ and further details are found in \cite{D+98}. A
semiclassical theory for the form factor in Hamiltonian ratchets, which will
be published elsewhere, requires to account for the simultaneous presence of
regular and chaotic regions in phase space. It relates the form factor
$K(x,t)$ to the respective contributions of these invariant sets to the
classical spatio-temporal distribution $P(x,t)$ (Fig.~4).

We benefitted from discussions with S.~Flach, P.~H{\"a}nggi, M.~Holthaus and
O.~Yevtushenko. MFO acknowledges financial support from the Volks\-wa\-gen
foundation.  TD thanks for the hospitality enjoyed during stays at the MPI
f\"ur Physik komplexer Systeme, Dresden, and the MPI f\"ur
Str\"omungsforschung, G\"ottingen, generously financed by the MPG.

\if\tc\else
\def\fw{150mm}\figI\newpage
\def\fw{150mm}\figII\newpage
\def\fw{150mm}\figIII\newpage
\def\fw{150mm}\figIV
\fi
\if\tc \end{multicols}\fi

\begin{thebibliography}{10}
\bibitem{FYZ-Parameters} $H=p^2/2+V(x,t)$ with
$\partial_{x}V(x,t)=(2\pi/\omega^2)\,[\cos(2\pi x)+0.6\cos(4\pi x+0.4)-2.3\sin(2\pi
t)-1.38\sin(4\pi t+0.7)]$ and $\omega=2.4$ corresponding to the parameter set
(3) of Fig.~1 in Ref.~\protect\cite{FYZ00} where we have scaled the spatial
and temporal period to 1.
\end{thebibliography}
\end{document}